\documentclass[11pt,twoside]{article}  
\usepackage[dvips]{graphicx}

\title{The Atacama Large Millimeter Array (ALMA)} 

\author{Alwyn Wootten\footnote{N. A. Project Scientist, ALMA} National Radio Astronomy Observatory\footnote{The National Radio Astronomy Observatory (NRAO) is a facility of the National Science Foundation operated under cooperative agreement by Associated Universities, Inc.},\\ 520 Edgemont Rd, Charlottesville,VA 22901 USA 
}

\begin{document} 
  \maketitle 

\begin{abstract}
The Atacama Large Millimeter Array, or ALMA,   is an international telescope project which will be built over the coming decade in 
Northern Chile. With over 7000 m$^2$ of collecting area comprised of 64
12m antennas arrayed over baselines up to 14 km in extent, ALMA will provide
images of unprecedented clarity and detail.  One revolutionary feature of
ALMA will be its ability to combine interferometric and single telescope
data, providing complete flux recovery.  ALMA will cover a spectral wavelength
range from 7mm to 0.3 mm or shorter wavelengths, providing astronomy with
its first detailed look at the structures which emit millimeter and submillimeter
photons,
the most abundant photons in the Universe.  

\end{abstract}



\section{INTRODUCTION}
\label{sect:intro}  
In this paper we review  ALMA goals, the chosen site at Chajnantor, the scope of the project, and a few of the technical hurdles and proposed solutions, as well as the ALMA schedule.

\subsection{Goals: Science Aims and Drivers}
\label{sect:goals}

Several publications$^{1,2}$ contain contributions describing the science goals
of ALMA.  In its basic formulation, the ALMA goal is to provide
images of unprecedented clarity and detail in the millimeter and submillimeter
spectral range.  This range contains two of the three primary peaks in the
electromagnetic spectrum of the Universe; these are the two containing
the preponderance of the observed energy in the Universe.
The largest of these is the peak from the 3
K blackbody radiation relic of the Big Bang.  That peak
occurs in the millimeter range of the spectrum, as expected for any
black body radiating at such a low temperature.   The
second strongest occurs at about 1.5 THz or 200 microns wavelength.  Light
of these wavelengths cannot penetrate the atmosphere, as it is absorbed
by water and other molecules--this maximum was identified
only recently through satellite observations.  Alas, with a satellite
one is limited as to the size of telescope with which one can observe and
anything we can currently put into space is far too small to give good
images of the energy sources comprising this second peak.  
From its characteristic
blackbody temperature of $\sim$20 K we are led to suspect that it is 
comprised of emission from the cold molecular clouds
from which stars and planets in the Universe have formed, and from 
the young galaxies full of dust which
host those molecular clouds.  Recently, some telescopes have made
progress in identifying the sources of this unknown radiation.  Much of
it--perhaps most of it--appears to come from tremendous episodes
of star formation in galaxies at the earliest stage of their creation. 
Unfortunately, many sources of submillimeter radiation have not been
identified optically, at least in part because their spectrum has been
wholly redshifted to wavelengths blocked by the terrestrial atmosphere and
in part because of tremendous amounts of dust endemic to the source, which
absorb optical light and re-emit it at longer wavelengths.
The only way to tell for sure is to get precise images of these sources. 
An instrument to provide these images must provide high resolution--matching
that which will be available at other wavelengths, or from 0.01'' to 0.1''
and it must provide the sensitivity to invest these images with high
dynamic range.  ALMA has been defined to achieve these goals.  It is
currently under construction by an international partnership.

With construction funding begun in FY2002, ALMA will be built over 
the coming decade in 
Northern Chile. ALMA will be a revolutionary telescope, operating over
the entirety of the millimeter
 and submillimeter wavelength band observable from the Earth's most lofty
regions. 

   \begin{figure}
   \begin{center}
   \begin{tabular}{c}
   \includegraphics[height=10cm]{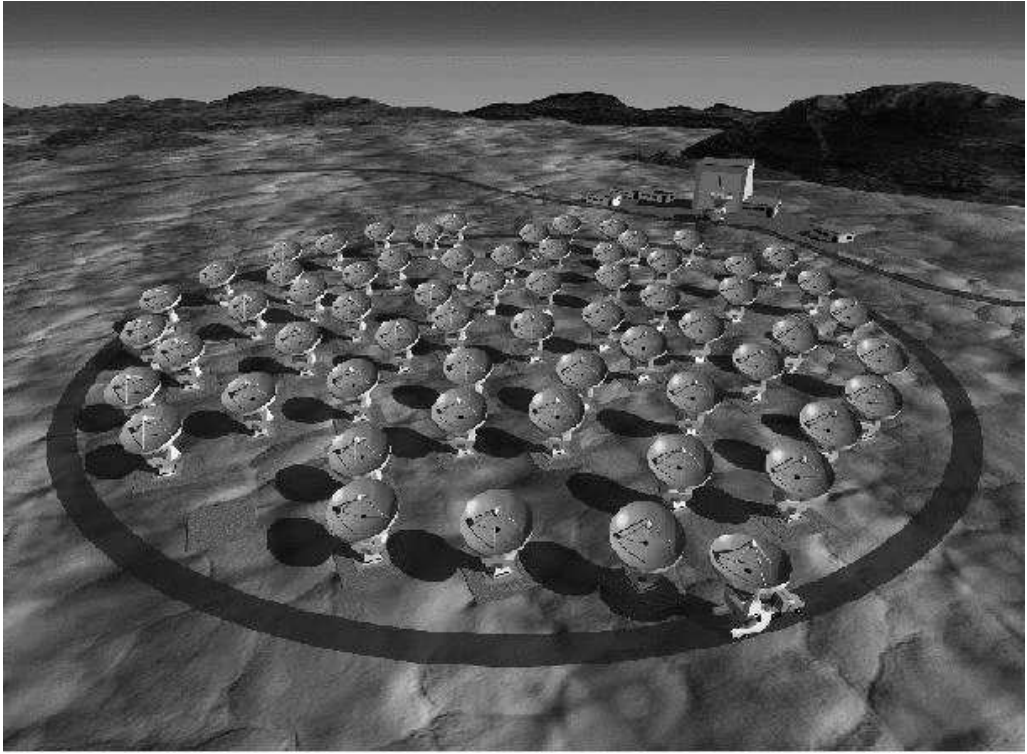}
   \end{tabular}
   \end{center}
   \caption[ Artist's rendition of ALMA in its compact configuration.  Image courtesy European Southern Observatory.] 
   { \label{fig:ALMA}  Artist's rendition of ALMA in its compact configuration.  Image courtesy European Southern Observatory.}
   \end{figure} 
Table 1 summarizes the specifications
for ALMA.
\begin{table}[h]
\caption{Summary of ALMA Specifications} 
\label{tab:spectable}
\begin{center}       
\begin{tabular}{|l|r|} 
\hline
\rule[-1ex]{0pt}{3.5ex}  Parameter & Specification \\
\hline
\rule[-1ex]{0pt}{3.5ex} Number of Antennas & 64  \\
\hline
\rule[-1ex]{0pt}{3.5ex} Antenna Diameter & 12m \\
\hline
\rule[-1ex]{0pt}{3.5ex} Antenna Surface Precision & $<$ 25 $\mu$m rss \\
\hline
\rule[-1ex]{0pt}{3.5ex} Antenna Pointing Accuracy & $<$ 0."6 rss \\
\hline
\rule[-1ex]{0pt}{3.5ex} Total Collecting Area & $>$7000 m$^2$ \\
\hline
\rule[-1ex]{0pt}{3.5ex} Angular Resolution & 0".02 $\lambda$ (mm) \\
\hline
\rule[-1ex]{0pt}{3.5ex} Configuration Extent & 150 m to 14 km\\
\hline 
\rule[-1ex]{0pt}{3.5ex} Correlator Bandwidth & 16 GHz per baseline\\
\hline 
\rule[-1ex]{0pt}{3.5ex} Spectral Channels & 4096 per window\\
\hline 
\rule[-1ex]{0pt}{3.5ex} Number of Spectral Windows & 8\\
\hline 
\end{tabular}
\end{center}
\end{table}
The ALMA specifications are described in more detail in the 
ALMA Construction Project Book, 
which like other project details may be found on the Worldwide Web at www.alma.nrao.edu. 
To fully enable the order of magnitude gain in resolution and two order
of magnitude gain in sensitivity over existing instruments which are ALMA's
goal, it will be located at an elevation of 5000m on the Chajnantor Alitplano
near San Pedro de Atacama
in northern Chile, as shown in Figure 1 and described below.
  Table 2 summarizes
the sensitivity of ALMA in various modes and at various frequencies, given the
atmosphere at the Chajnantor site.  The calculations assume 1mm of preciptable
water vapor above the site, observation at 1.3 air masses (40$^o$ zenith
angle) and antenna performance as given in Table 1.

\begin{table}[h]
\caption{Summary of ALMA Sensitivities} 
\label{tab:senstable}
\begin{center}       
\begin{tabular}{|l|c|c|r|} 
\hline
\rule[-1ex]{0pt}{3.5ex} Band no. & Frequency & Continuum & Line (1 km s$^{-1}$) \\
           & Range (GHz)$^a$  & (mJy; 60s) &(mJy; 60s) \\
\hline
\rule[-1ex]{0pt}{3.5ex}1 & 31.3--45 & 0.02 & 5.1 \\
\hline
\rule[-1ex]{0pt}{3.5ex}2 & 67--90 & 0.028&4.9  \\
\hline
\rule[-1ex]{0pt}{3.5ex}3$^b$ & 84--116 & 0.027 & 4.4 \\
\hline
\rule[-1ex]{0pt}{3.5ex}4 &125--163 & 0.039 & 5.1 \\
\hline
\rule[-1ex]{0pt}{3.5ex}5 &163--211 & 0.52 &5.9  \\
\hline
\rule[-1ex]{0pt}{3.5ex}6$^b$ &211--275 & 0.071 & 7.2 \\
\hline
\rule[-1ex]{0pt}{3.5ex}7$^b$ & 275--370 & 0.120 & 10 \\
\hline
\rule[-1ex]{0pt}{3.5ex}8 & 385--500 & 0.34 &26  \\
\hline
\rule[-1ex]{0pt}{3.5ex}9$^b$ & 602--720 & 0.849 & 51 \\
\hline
\rule[-1ex]{0pt}{3.5ex}10 & 787--950 & 1.26 & 66 \\
\hline
\end{tabular}
\end{center}
\vspace{0.3cm}
$^a$ Frequency ranges as given in {\it Specifications for the ALMA
Front End Assembly}, AEC Version 1, 18 October, 2000.\\
$^b$ These  4 bands are the highest priority bands, the others
are the second priority bands, as judged by the ALMA Scientific Advisory
Committee.\\
\end{table}
\subsection{ALMA Partners} 
\label{sect:partners}
ALMA has gained widespread support as an instrument astronomy needs
to develop astronomical ideas through observation in the twenty-first century.
The project, previously known at the Millimeter Array (MMA), 
has been endorsed by both the 1991 and
2000 Astronomy and Astrophysics Survey Committees of the National
Academy of Sciences (US) as among the highest priority items for astronomical
facilities to be constructed.  In France, at a colloquium at Arcachon in
March 1998 organized by CNRS/INSU, French astronomers put construction of ALMA,
in its interim form as the Large Southern Array (LSA),
as highest priority for future instruments.
In Canada, ALMA was identified as top priority by the National Research 
Council's Long Range Planning Panel on Canada's Future in Astronomy until 2015.
Dutch astronomers likewise established ALMA as the top priority for
instrumentation in the coming decade.  
In the U.K., the Astronomy Vision Panel
of PPARC identified ALMA as the highest priority medium-term project.

This large set of astronomical community endorsements has gained ALMA a number
of international partners.
North American ALMA partner institutions include
the United States National Science Foundation, through its National Radio
Astronomy Observatory facility operated by Associated Universities, Inc., and the Canadian 
National Research Council.  NRAO works in cooperation
with a University consortium including the Owens Valley Radio Observatory 
of Caltech and the 
Berkeley Illinois Maryland Association.  European partners include the
European Southern Observatory;
The Centre National de la Recherche Scientifique (CNRS), France;
The Max Planck Gesellschaft (MPG), Germany; The Netherlands Foundation for 
Research in Astronomy, (NFRA); Nederlandse Onderzoekschool Voor Astronomie, (NOVA);
The United Kingdom Particle Physics and Astronomy Research Council, (PPARC);
The Swedish Natural Science Research Council, (NFR); and the Oficina de
Ciencia y Tecnologia and Instituto Geografico Nacional (IGN), Spain.
 Chile, as host
nation for the ALMA project, participates through its presence on the
ALMA Coordinating Committee, the ALMA Science Advisory Committee and by providing the excellent site high in the Andean Altiplano and support for it.
The National Astronomical Observatory of Japan may join the ALMA 
consortium soon.

\subsection{Location: the Site at Chajnantor, Chile} 
\label{sect:site}
In May 1998 NRAO recommended construction of the MMA on a site in Region II
of northern
Chile which lies on a high plain at the foot of three ancient 
volcanic peaks, Cerro Toco,
Cerro Chajnantor and Cerro Chascon.  The site 
(longitude 67d 45m W, latitude -23$^o$ 01' S)lies near the Tropic of
Capricorn,  about 50 km east of the 
historic village of San Pedro de Atacama, 130 km southeast of the mining
town of Calama, and about 275 km ENE of the coastal port of Antofagasta. 
It lies close to the border with Argentina and Bolivia beside the paved Paso
de Jama road into Argentina and gas pipelines connecting Argentina sources
with Chilean mines.  The mean elevation is about 5000m (16400 ft).
Several sites had been tested but
as scientific interest increased for the highest frequencies, sites at the
highest altitudes became favored.  Testing of the Chajnantor site began
in April 1995 and continues to the present, a joint effort of NRAO and
the European Southern Observatory (ESO).  The Nobeyama Radio Observatory
(NRO) operates a similar testing facility nearby at Pampa la Bola.
The testing operations continue with the involvement of the Chilean
university community.  The land is administered by the Chilean Ministry
of National Assets, having been set aside as a protected region for science
by Presidential decree.

Up-to-date details of the monitoring of a number of parameters critical
to ALMA's success continue and current details from the instrumentation
may be found at the ALMA web site.  Some salient characteristics include:
the median annual temperature is -2.5$^o$C with annual 50th percentile
winds of 10.4 m s$^{-1}$.  The average barometric pressure is only 55 percent
of the value at sea level.  Humidity averages 39\% and ultraviolet
radiation is about 170\% that at sea level.  Transparency at 225 GHz
has been monitored for several years; the 50th percentile zenith optical
depth at this frequency is
0.061 corresponding to a column of precipitable water of a little more than
1mm$^3$.  With such a low water column, observations are possible at the
atmospheric windows covered by ALMA Bands 9 and 10 (see Table 2) for
roughly half of the time.  Direct observations of atmospheric transparency
including the supraterahertz windows at 1.035, 1.3 and 1.5 THz have been
published$^4$ showing transmission of up to 20\% in the highest frequency
window.

\subsection{Science} 
\label{sect:science}

The mission of ALMA is to produce detailed spectral line and continuum
images of objects emitting radiation in the millimeter and submillimeter
spectrum accessible from the surface of the Earth.  These high dynamic
range images, covering a wealth of spatial scales and featuring total
flux recovery,  
will allow ALMA scientists to study the
formation of galaxies, stars and planets and the distribution of the
chemical precursors necessary for life.  Specifically, ALMA will allow
astronomers to address a number of topics of high interest.

As mentioned above, the photons comprising the visible/infrared and
submillimeter portions of the spectrum arise from two distinct features
in the overall spectrum of background radiation.  Images of the sky
in these two bands are therefore complementary.
The distinct nature of the submillimeter sky far from the Galactic Plane
is compellingly illustrated by 
comparing the Hubble Deep Field (HDF) imaged at optical or infrared
wavelengths  with a submillimeter view (SCUBA 850$\mu$m $^5$).  
As redshift increases, the volume
of the Universe surveyed increases rapidly; this fact 
combined with an increasingly bright spectrum being brought into the observing
band by higher redshifts (the so-called K-correction--$S_\nu \propto \nu^{3-4}$)
ensures that the brightest sources at submillimeter wavelengths are 
distant (z$>$1) dusty galaxies.  Furthermore, a generally higher star 
formation rate in the
earlier epochs$^6$ adds to the dominance of high redshift
objects in the submillimeter.
Comparison of the expected blackbody spectrum of the ultraluminous 
starburst galaxy Arp 220 as seen at high redshifts with 
the frequency coverage and the sensitivity of ALMA suggests that The
Atacama Large Millimeter Array should be able to detect Arp 220-like dusty 
starburst galaxies out to a redshift of 10 or more.  Galaxies like
the present day Milky Way can also be detected out to z beyond 3.

Although continuum emission from dust holds great interest, ALMA will also 
provide spectral line images to detail the kinematics of the gas, its excitation
and its chemical and isotopic composition throughout the history of the
universe.  In the submillimeter and millimeter bands, the gas and dust
enveloping 
galactic nuclei may be imaged without the optical obscuration which affects
optical or infrared observations; the kinematics may be measured on spatial
scales less than 100 pc or so.  Various redshifted
spectral lines--for example, all rotational
transitions of CO and infrared fine structure lines such
as [C II], [C I], [N II] and [O I]--can be detected and studied to derive
the molecular gas content and elucidate the character of 
star forming activity.  

Millimeter/submillimeter astronomy has discovered 
most of the $>$120 known interstellar molecules.  These molecules may be
quite complex--up to 12 atoms have been found in those small dense regions
whose imaging will be ALMA's strength.  ALMA offers complete coverage of 
the available windows from a site of unprecedented atmospheric clarity.
ALMA improves currently available spectral sensitivity by more than 
an order of magnitude (the 
weakest lines detected today are of order 1 Jy km/s in strength; compare to 
the ALMA sensitivity in Table 2 above).  But as sensitivity improves,
understanding of an already crowded spectrum may not be possible without
an accompanying increase in spatial resolution.  This
ALMA also improves, by nearly two orders of magnitude, reaching 
resolutions of ~10 milliarcsec.  Furthermore, the throughput, measured by the
spectral bandwidth open to ALMA (16 GHz), exceeds by several times that 
available now.  A recent spectral survey from the 12m telescope revealed
some 15,000 lines in one spectral window toward the seven sources covered
(Turner, private communication); ALMA's potential is clearly tremendous.

As stars form from a mixture of these molecules and dust, disks remain to form 
planets.  Most T Tau stars possess disks with masses from 0.1 to 0.001 solar 
masses and sizes of 100-500 AU. Similar masses and sizes of disks appear to 
be present in protostars, embedded in dense molecular cloud cores.  As the 
disks evolve, the chemistry continues to evolve, adapting to local 
conditions.  With resolution reaching 5AU 
in nearby (140 pc) systems, ALMA will observe 
the chemical makeup of the disks as a function of vertical height (the disk 
`atmosphere', with ultraviolet processing important higher up and 
condensation onto icy dust grains important mid-plane) and of radial 
distance from the star.  Jets driven by the young star 
generate shocks into a cocoon of material surrounding the jet, chemically 
processing that material.  Dutrey$^7$ has summarized ALMA's contribution
to molecular disk studies; many molecules have been observed 
in the disks about T Tau stars such as DM Tau and GG Tau, including CN, HCN, 
HNC, CS, HCO+, C$_2$H and H$_2$CO, suggesting a photon-dominated chemistry; 
even at 
4'' (600 AU) resolution changes in chemical structure of disks may be noted 
(e.g. LkCa15; Qi$^8$).  With ALMA, these structures will be 
more sensitively imaged; even time-resolved changes should be distinguished.  
ALMA will provide the sensitivity and resolution necessary to probe the 
planet-forming disk midplane.

Fairly complex molecules have been detected in the interstellar medium.  
The simple sugar glycolaldehyde, a building block for 
more complex sugars ribose and glucose, was recently detected.  Detection of 
its chemically reduced form, the ten atom molecule ethylene glycol, has just 
been announced.  Searches for the simplest amino acid, the ten atom molecule 
glycine
(NH$_2$CH$_2$COOH) have proven negative, lacking sensitivity.  Glycine has a 
complex spectrum, exhibiting about a dozen lines per GHz on average.  The 
Green Bank Telescope offers the sensitivity and wide bandwidth for a renewed 
search; ALMA will add to these attributes directivity, reducing spectral 
confusion in the complex regions where large molecules congregate.  Detection 
may also be possible of other prebiotic molecules, such as adenine or uracile.

Molecules formed in dense regions may become incorporated in subplanetary or 
planetary objects.  In cold regions, dense gas becomes highly deuterated.  
Elevated deuterium levels have been observed in some disks, as well as in 
cometary comae, suggesting some dense gas maintains its identity as planetary 
systems form. Evidence from the solar system suggests that some organics 
remain, even in the inner solar system, as they have become incorporated into 
discrete bodies.  Carbonate blebs in the ALH84001 Martian meteorite maintained 
a heterogeneous pattern of magnetization suggesting transfer of the meteorite 
from Mars to Earth did not result in heating above 40$^o$C, 
Even after incorporation into larger bodies,
organics may be transferred between bodies.  Hence
a direct connection may exist between complex molecules observable by ALMA
in interstellar space and the complex prebiotic molecules found in small
condensed objects in the solar system.

In solar system objects, ALMA will obtain unobscured subarcsecond images.
As an example, the flow of sulfur oxides from the volcanoes on Io may be
imaged to differentiate their origins.

ALMA will be able to detect wobbling motion in stars caused by
planets orbiting them.  ALMA will be able to detect planets in the process
of formation.  ALMA will only marginally detect thermal emission 
from nearby extrasolar large planets.
  Although it will not be able to image even nearby Earthlike
planets, it will be able to detect the interstellar chemicals which are 
available for the nourishment of early life.

As stars evolve, nuclear processes in their interiors results in a
striated structure.  As their energy sources change according to local
physical conditions, the star must adjust hydrodynamically, often shedding
shells of processed material into the interstellar medium which cool and
form dust and molecules.  ALMA will image these shells to reveal the
isotopic and chemical gradients 
that reflect the chronology of invisible stellar nuclear processing.

ALMA is the only instrument, existing or planned which provides the
combination of sensitivity, angular resolution and frequency coverage
to achieve these scientific goals.  By
providing high fidelity images at subarcsecond resolution
 ALMA will complement both the present
generation of gound-based telescopes, such as VLT, Gemini and Keck and
their successors, GSMT or CELT, and space telescopes such as the HST and NGST.
ALMA's images of cool thermal emission complements optical/infrared
data provided by these telescopes to enable astronomers to explore
all of the peaks of the electromagnetic spectrum at similar resolution and
high sensitivity.

\section{ Technical challenges and proposed solutions, schedule}
\label{sect:intro}

\subsection{ The Antenna: Mechanical Engineering at the Heart of the Array
 }
\label{sect:antenna}

Each of the two major partners plus Japan have contracted for construction 
of a 12m diameter prototype antenna meeting the specifications 
imposed by the scientific goals for ALMA.
These prototype ALMA antennas will be delivered to
The ALMA Test Facility at
the NRAO's Very Large Array (VLA) site near Socorro, N.M., for tests to
ensure it meets the very demanding specifications of this project. 
VertexRSI of Santa Clara, California is the U.S. contractor for one
antenna; this antenna is currently being assembled for tests to begin
in November 2002.    A second prototype design will be delivered to the same 
site in the spring of 2003 by a European consortium including Alcatel 
and European Industrial Engineering.
These two antennas will be tested in order to determine which design is
most appropriate for ALMA; both designs are the property of the ALMA project.
 The successful design will become the production design, 
to be built by the winner of a bidding process.
A third prototype, built under contract with Mitsubishi,
is expected to be delivered through an arrangement with the 
National Astronomical Observatories of Japan in 2003.
   \begin{figure}
   \begin{center}
   \begin{tabular}{c}
   \includegraphics[height=15cm]{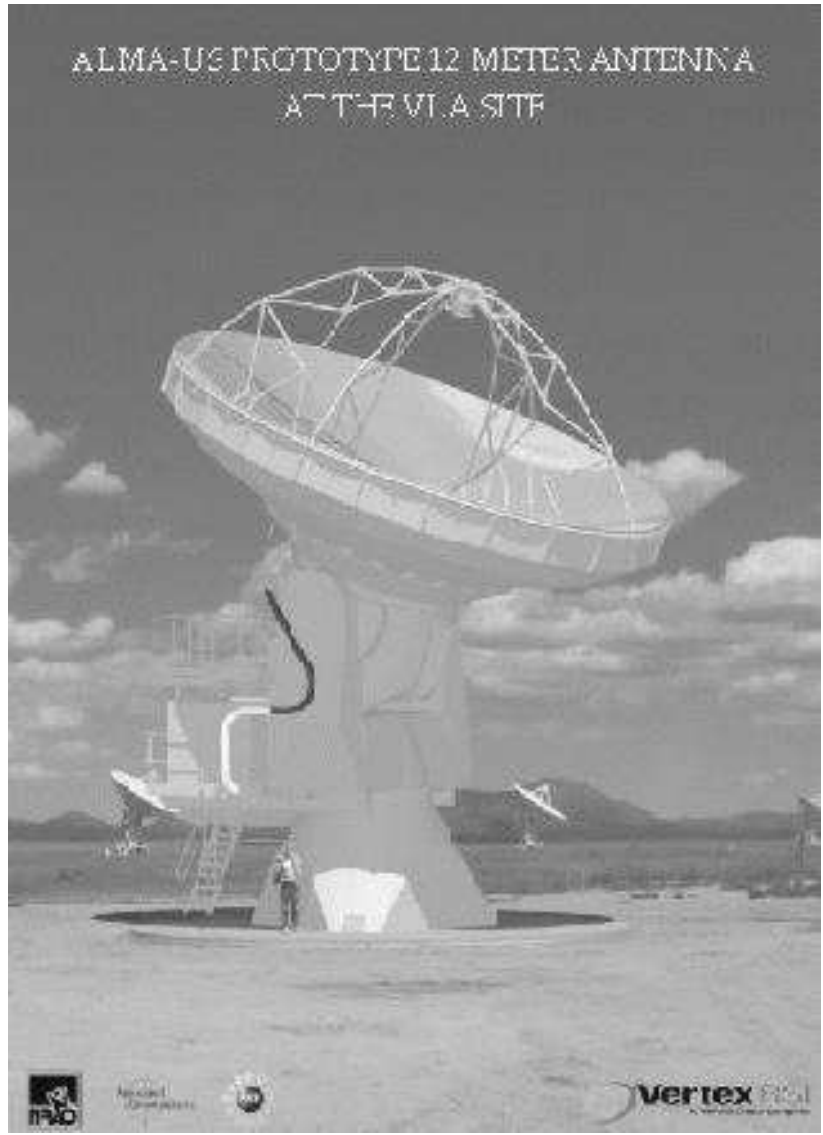}
   \end{tabular}
   \end{center}
   \caption[ Artist's rendition of ALMA in its compact configuration.  Image courtesy European Southern Observatory.] 
   { \label{fig:Vertex}   An engineering rendering of the 12 meter diameter prototype antenna that is being fabricated by VertexRSI to the very demanding ALMA specifications.  The prototype antenna is under construction at the NRAO Very Large Array site for testing, as shown in this drawing.}
   \end{figure} 

The ALMA prototype antenna makes extensive use of carbon fiber reinforced 
plastic (CFRP) technology in order for the antenna to maintain
a stable parabolic shape in the harsh thermal and wind environment 
characteristic of the ALMA site at the 16,500 feet.  The antenna surface 
accuracy must be better than 25 microns
to enable efficient observations at the very highest frequency.  Much of 
the time, ALMA will image fields larger than the primary beam, which means 
that multiple pointings of the antenna will be fed into a single image which
will include interferometric and total power data alike.  For accurate
imaging, the antenna and its calibration must be quite stable.  For example,  
this requires that the antennas maintain a pointing accuracy better than 0.6 
arcseconds, despite breezy conditions (50 percentile 6.5 m s${-1}$).  The 
Chajnantor site affords no vegetative cover of consequence, so windblown 
grit and dust will occur and must not degrade the performance of the antenna.  
At such an altitude, temperature extremes occur--the annual median 
temperature is below freezing and can range $\pm$ 20 C--yet antenna 
performance must be maintained.  The ultraviolet 
radiation at this altitude is 170\% of that at sea level, requiring 
construction of material robust to UV damage.  All these factors 
provide a challenge to modern antenna design.

As the array is reconfigurable, the antennas must be transportable on an 
occasional basis by a transporter.  The maximum baselines available are on 
the order of 14km.  Current plans are to move antennas on a few-per-day basis 
along a self-similar configuration of roughly spiral geometry out to the 
largest configuration, where topography heavily constrains the design.  
This will provide images with a range of detail and a resolution of
0.015'' $\lambda _{mm}$.

\subsection{ The receivers:  The world's largest superconducing array. }
\label{sect:receivers}

 Many laboratories worldwide contribute to the ALMA receiver effort.  The
receivers will cover the entire observable submillimeter/millimeter spectrum 
observable from ALMA's superb site.  Eventually, ten receiver 
bands will span this region.   Currently, receivers for the lowest two bands 
are planned to be transistor amplifiers.  For frequencies above 
$\sim$80 GHz the receiver designs
employ superconducting tunnel junction mixers cooled to below 4K with 
transistor 
amplifiers operating over a 4-12 GHz or 4-8 GHz range to provide the correlator
with 16 GHz of data.  
During initial construction, four receiver bands will be
built and deployed as identified in Table 2.
Orthogonal polarizations will be received, bringing ALMA's eventual 
total to over one 
thousand receivers--the most extensive superconducting electronic receiving 
system in astronomy.   The complement is augmented by a 
system of radiometers operating at 183 GHz which will monitor water vapor 
in the atmosphere above each telescope for path length correction of the 
incoming signals in a
fashion analogous to the use of adaptive optics used on telescopes
at visual and infrared wavelengths.

To cover the observable spectrum in ten bands, radiofrequency bandwidths up 
to 30\% will need to be covered by receivers at some bands.  Below 350 GHz, 
the receivers will be constructed to operate in single sideband  to lessen 
atmospheric noise contributions. 

To achieve this performance, the receivers will be housed in cartridges of a 
modular design,with all ten bands enclosed within a dewar of roughly 1m 
diameter and stages cooled to 70, 15 and 4K.  For each band, a modular 
cartridge will be developed which fits into the dewar from the bottom and 
which is held in place by flexible thermal links.  Thus, all bands share the 
focal plane, and optics and calibration devices will sit atop the dewar.

\subsection{ The Correlator: Achieving 1.6 x 10$^{16}$ multiply/add operations 
per second }
\label{sect:receivers}

The ALMA Correlator will be located at the Chajnantor site.  The analog input 
from the receivers on 64 antennas--8 spectral windows, 
each of 2GHz bandwidth--will be 
digitized and transmitted at a rate of over 100 Gigabits per second from each 
antenna.  The signal travels over fiber optic cables to digital filters, 
then to the correlator.  The correlator must achieve 1.6 $\times$ 10$^{16}$ multiply 
and add operations per second.
It cross-correlates signals from 32$\times$63=2016 pairs of antennas on 16 msec 
timescales; it also autocorrelates signals from 64 antennas on 1 msec 
timescales, producing 32 Gbyte s$^{-1}$ output.
The correlator is currently under construction at the Central Development 
Laboratory of the NRAO on the University of Virginia grounds.  Its design 
offers flexibility of selection of bandwidth,
spectral window placement and a reasonably low power requirement of 150 kW.

Correlator design and construction require a particularly long lead time; 
given the pace of technology prudence requires that ALMA consider the next 
generation correlator as the current generation correlator is constructed.  
Therefore, ALMA has fostered the design of the 'G2X Correlator' for possible 
deployment in the next decade.  That design is currently scoped to provide 
twice as many channels in high dispersion mode as the ALMA Correlator; more 
than an order of magnitude more in low dispersion mode.  
Furthermore, the G2X Correlator concept improved sensitivity through three-bit 
digitization in all modes rather than in the high dispersion modes only as 
provided by the ALMA Correlator.  The concept envisions twice as many spectral 
windows (16) as the ALMA Correlator.  

\subsection{  Schedule }
\label{sect:schedule}
The current ALMA Schedule is based upon the European construction phase 
beginning in mid-2002.  The first antennas would then arrive in Chile sometime 
in late 2005.  Commissioning observations would then begin late in 2006 and 
early science operations late in 2007.  
Under this schedule construction would be 
finished at the end of 2011 with full operations commencing in 2012.
\subsection{ References}
{\small

 Alwyn Wootten~\& F. Schwab\ {\it Science with a Millimeter Array} 
NRAO Workshop Vol. No. 14, Charlottesville: National Radio Astronomy 
Observatory (NRAO), edited by A. Wootten and F. R. Schwab, National
Radio Astronomy Observatory, Green Bank, W. Va., 1988.

 Alwyn Wootten~\ {\it Science with the Atacama Large Millimeter Array},
ASP Conf. Ser. 235, edited by A. Wootten, Astronomical Society of the Pacific,
San Francisco, 2001.

S. Radford~\& R. Chamberlin ``Atmospheric Transparency at 225 GHz 
over Chajnantor, Mauna Kea, and the South Pole'' ALMA Memo 334, 2000.

 S. Matsushita, H. Matsuo, J. R. Pardo \& S. J. E. Radford
``FTS Measurements of Submillimeter-Wave Atmospheric Opacity at Pampa la Bola 
II: Supra-Terahertz Windows and Model Fitting''
{\it Publ. Ast. Soc. Japan},  {\bf51} pp. 603-610, 1999.

 Hughes, D. et al., ``High-redshift star formation in the Hubble Deep 
Field revealed by a submillimetre-wavelength survey'', {\it Nature}, {\bf394} pp. 241-7, 1998.

 Blain, A.\ W., Smail, I., Ivison, R.\ J.\ \& Kneib, J,, ``The history of star formation in dusty galaxies'',  {\it MNRAS}, {\bf302} pp. 632-648, 1999.

 Dutrey, A. in {\it Science with the Atacama Large Millimeter Array},
ASP Conf. Ser. 235, edited by A. Wootten, Astronomical Society of the Pacific,
San Francisco, pp. 215-224, 2001.

  Qi, Chunhua, {\it Aperture synthesis studies of the chemical 
composition of protoplanetary disks and comets}, Ph. D. Thesis, California
Institute of Technology, 2001.
}



\end{document}